# Unpacking Interaction Profiles and Strategies in Human-AI Collaborative Problem Solving: A Cognitive Distribution and Regulation Perspective

Zhanxin Hao[1], Xiaobo Liu[1], Jiaxin Fan[1], Yun Long[1], Jifan Yu[1], Wenli Chen[2], Yu Zhang[1*]

1 School of Education, Tsinghua University
2 National Institute of Education, Nanyang Technological University

## Abstract

This study adopts an integrated distributed cognition and co-regulation perspective to examine the collaboration patterns and dynamics of human-AI collaboration when college students collaborating with AI for complex problem-solving. Through cluster analysis, three distinct collaborative problem-solving modes were identified in this study: Delegated Reasoning (DR), Concerted Interpretation (CI), and Delegated Elaboration (DE). This study found that the DR group achieved the highest task performance, significantly outperforming the CI group. Additionally, the semantic similarity between human and AI discourse was notably the highest in the DR group. In contrast, the CI group reported significantly greater use of self-regulation strategies. These findings uncover a critical tension between the efficiency of the distributed system and the depth of human learners' regulatory engagement. Insights from this study offer valuable implications for the future design of AI-empowered educational tools and student-AI collaborative learning frameworks.

[1] This work was supported by the Beijing Educational Science Foundation of the Fourteenth 5-year Planning (BAEA24024).

Corresponding details: zhangyu2011@tsinghua.edu.cn

# 1. Introduction

Collaborative problem solving (CPS) is becoming increasingly important in an era when complex tasks are no longer addressed only through human-human collaboration, but also through interaction with increasingly capable AI agents. Generative AI systems can now contribute to CPS by analyzing goals, synthesizing information, generating alternatives, and supporting iterative refinement across multiple modalities (Kim et al., 2022). More broadly, contemporary technological discourse has even envisioned scenarios in which a single individual collaborates with multiple AI agents to build a billion-dollar company, a phenomenon described as the "one-person unicorn" (The Economist, 2025). Although such visions remain speculative, they vividly illustrate the expanding role of human-AI collaboration in complex task environments. Against this backdrop, CPS has become an increasingly important context for understanding how humans and AI agents jointly solve problems through shared understanding, coordinated effort, and adaptive interaction.

As generative AI fundamentally operates based on its underlying algorithms and is highly reactive to user input, its collaborative effectiveness largely depends on how the user employs it. Existing research reveals that students' varying prompt literacy influences the quality of AI agents' responses (Lo, 2023; Kim et al., 2025). More fundamentally, CPS involves not only social collaboration, such as communication and role negotiation, but also cognitive coordination, such as exchanging feedback and iteratively refining solutions. What matters, therefore, is not merely whether and how students use AI, but how cognitive work is distributed and regulated between human and AI collaborators during problem-solving processes. According to the theory of Distributed Cognition, cognitive processes are not confined to one person's mind but are distributed across people, artifacts, tools, and time (Hutchins, 2000). This perspective suggests that the ways students allocate, offload, and coordinate cognitive activities with AI agents may be central to problem-solving success. However, existing research has paid limited attention to how learners distribute cognition with AI agents in CPS, how such distribution is monitored and adjusted, and which patterns of human-AI collaboration are associated with more effective learning outcomes.

To address this gap, the present study examines the interaction profiles that emerge when college students collaborate with AI agents in problem-solving tasks.

Through fine-grained analysis of dialogic interactions, the study identifies distinct patterns of human-AI collaboration, investigates how these patterns differ in the distribution of cognitive work between students and AI agents, and examines which patterns are associated with stronger regulatory engagement and task performance. In doing so, the study extends distributed cognition research into AI-supported collaborative learning contexts and provides empirical evidence on the effectiveness of specific human-AI cognitive distribution patterns. This study proposes the following research questions (RQs):

RQ1. How do the communicative functions and cognitive responsibilities differ between student and AI agents during CPS?

RQ2. What distinct interaction patterns emerge when students collaborate with AI agents in CPS?

RQ3. How do different student-AI interaction patterns relate to students' regulation and task performance?

## 2. Literature Review

### 2.1 Theoretical framework

The current study adopts the theory of distributed cognition (Hutchins, 1995; Perkins, 1993; Nardi & O'Day, 2000) to understand the complex interactions among human students and AI agents. Hutchins (1995, 2000) introduced the concept of distributed cognition through studies of ship navigation and cockpit operations, arguing that cognition is distributed across multiple processing units, including both human and non-human components. This perspective shifts the focus from isolated cognitive processes to a broader system of interdependent elements working together to accomplish goals.

The theory of distributed cognition posits three analyzable mechanisms: 1) representation flow, the ways information is encoded and transformed across artifacts/agents; 2) coordination, the routines that synchronize contributions; and 3) cognitive offloading/scaffolding, the relocation of processing to external structures (Hutchins, 1995; Hollan et al., 2000; Kirsh, 2010). Traditional CPS research has used this framework to understand how human teams distributing labor and expertise (e.g.,

Janssen et al., 2012), suggesting that successful CPS depends on the effective coordination of group-wide cognition: teams must establish common ground, manage shared representations, and leverage both individual ideas and collaborative supports (Daradoumis & Marques, 2002; Xu et al., 2023). Some empirical studies have examined the collaborative processes in various problem-solving scenarios, such as group programming task (Deitrick et al., 2015) and classroom activities (Xu et al., 2023). However, distributed cognition research in human-AI collaboration contexts remains limited.

The advent of generative AI brings new opportunities and challenges in examining the distributed cognition mechanism: unlike static tools in Hutchins' (1995) classic models, contemporary AI dynamically assumes roles ranging from co-constructor of knowledge to adaptive scaffolder, structuring how cognition is distributed in learning contexts. This study extends the research in cognitive distribution in CPS from interpersonal perspective to human-AI perspective, with the aim to optimise evidence informed design of human-AI collaboration to augment human learning in AI-supported learning environments.

In AI supported CPS, both students and AI agents not merely engage in cognitive coordination but also in active adjustment of the collaboration process. While distributed cognition theory explains how cognitive tasks are shared and distributed across a system, how this distribution is managed and regulated was under studied. To address the dynamics of human-AI CPS, this study integrates distributed cognition theory and regulation of learning theories including self-regulation, co-regulation and socially-shared regulation.

Self-regulation has traditionally been conceptualized as an active, constructive process whereby learners set goals for their learning and then attempt to monitor, regulate, and control their cognition, motivation, and behavior (Pintrich, 2000; Zimmerman, 2000). Although this framework has been widely applied in educational research (e.g., Panadero, 2017; Sitzmann & Ely, 2011), it predominantly focuses on individual learning processes. From a socio-cognitive perspective, learning is inherently situated within a social environment where interactions with peers and

teachers—and by extension, AI agents—significantly influence individual cognition (Bandura, 1986; Järvelä et al., 2010). Successful collaborative learning requires more than individual regulation; it demands the coordination of regulatory processes among group members (Rogat & Linnenbrink-Garcia, 2011). Researchers have expanded the traditional self-regulation framework to social regulation (Hadwin & Järvelä, 2011), specifically, co-regulation and socially-shared-regulation. According to Järvenoja et al. (2013), co-regulation is an interactive process where group members support or influence each other's individual regulatory efforts (often involving scaffolding from a peer or teacher). This is distinct from socially-shared-regulation, which represents a symmetrical and genuinely shared process where multiple members jointly regulate their collective activity to achieve a common goal (Vauras et al., 2003).

As shown in **Figure 1**, we propose a more integrated framework for AI-supported collaborative problem solving, one that explains both how cognitive work is distributed and how that distribution is dynamically regulated by human and AI agents.

**Figure 1**
*Human-AI Cognitive and Regulatory Distribution (HA-CORD) Model*

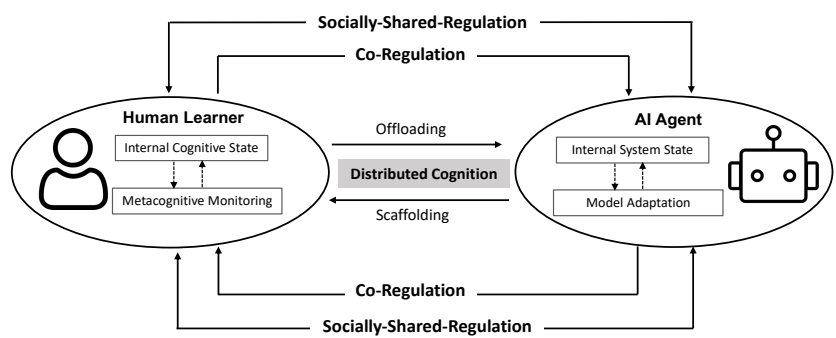

*2.2 Student-AI collaboration*

Student-AI collaboration refers to students and AI systems working together on learning tasks by leveraging their respective strengths and compensating for each other's weaknesses, thereby achieving outcomes that surpass the independent performance of either a human learner or a single AI system (Ouyang & Jiao, 2021;

Dellermann et al., 2019). A growing body of research has begun to examine the student-AI collaboration patterns in problem-solving contexts, identifying diverse modes of student–AI collaboration from different analytical lens. Focusing on the role of student as dominance or leader, Ouyang and Jiao (2021) synthesized prior literature and proposed three paradigms of student–AI relationships: 1) AI-directed, learner-as-recipient, 2) AI-supported, learner-as-collaborator, and 3) AI-empowered, learner-as-leader. Similarly, Zhu et al. (2024) proposed a related tripartite classification—even contribution, human-leads, and AI-leads—and found that most students in their study (77.21%) perceived themselves as either leading the collaborative problem-solving process or contributing equally with AI. From the perspective of how students process AI-provided information, Amoozadeh et al. (2024) categorized three types of student responses to AI outputs: accepting the entire response, selectively using parts of it, and rejecting it to retry. While these typologies offer valuable insights, such studies often rely heavily on subjective summarization, with relatively limited in-depth exploration of the data and insufficient engagement with established theoretical frameworks.

At a more granular level of cognitive processing, the literature converges on the finding that AI can effectively undertake lower-order cognitive processing and, in some cases, parts of complex reasoning, thereby providing cognitive augmentation that preserves human cognitive resources for higher-level strategic decision-making. For example, Wang et al. (2025) found that generative AI used in programming education can provide conceptual elaboration and error diagnosis, reducing learners' cognitive load by 32%. In addition, AI can function as a metacognitive tool by prompting reflective questions to support the planning and monitoring phases of learning, thereby facilitating socially shared regulation among learners (Edwards et al., 2025). However, several studies also reveal that student–AI collaboration often remains low-level and instrumental in nature. Tong et al. (2025) observed that high school students collaborating with ChatGPT-4o to solve scientific problems frequently engaged in a "pose–receive" pattern—inputting a query (often via screenshots) and directly accepting the output—rather than initiating high-quality, inquiry-oriented dialogue. Students tended to treat ChatGPT-4o as a static answer-retrieval tool, often accepting incomplete or flawed answers without further probing or critique. Misiejuk et al. (2025) found that early-stage course learners and lower achievers often relied on simple prompt–response interactions, whereas higher-achieving students and those in later stages of a course were more likely to embed contextual and normative information into their initial prompts, explicitly guiding AI output generation, and iteratively refining responses through multi-turn prompting.

Existing research on human-AI collaboration and interaction has largely focused on broad observations of online behavior instead of more systematic analysis of interaction dynamics—that is, how collaboration unfolds turn-by-turn and how students regulate and adapt their strategies in response to AI output at different stages of problem solving. In addition, these studies seldom adopted well-established theoretical frameworks when examining the human-AI dynamics.

There is limited empirical evidence revealing the relationship between student-AI collaboration patterns and learning outcomes, particularly in terms of cognitive outcomes, metacognitive engagement, and learning experience. Addressing these gaps requires moving beyond post-hoc categorization toward process-oriented, theory-informed coding that can reveal functional relationships in student–AI collaboration.

To address the research gaps, the present study adopts a fine-grained dialogue coding approach grounded in established collaborative learning theories, in particular, collaborative problem-solving theory, distributed cognition theory, and socially-shared regulation theory could provide a structured lens for understanding. By systematically analyzing each turn of student–AI interaction, this study aims to identify distinct human-AI collaboration dynamics, examine how cognitive responsibilities are distributed between human and AI partners, and generate empirically findings to inform the design of AI systems that foster effective and meaningful student–AI partnerships.

3. Methods

*3.1 Design of Multi-Agent AI System*

To simulate students' collaborative problem-solving behaviors in authentic scenarios as closely as possible, this study designed four distinct collaborative AI agents with varying personalities: Student M, Student J, Student H, and Student Q. These AI agents were differentiated primarily through tailored prompt engineering, which assigned unique behavioral and cognitive traits to each, as shown in Figure 1. These agents were deployed on an online learning platform ([www.maic.chat](www.maic.chat)) to interact with students (Yu et al., 2025). We operated these agents as a computational ensemble where a director agent dynamically managed agent participation through contextual analysis. Conceptualized as directing a theatrical performance, the director agent evaluated real-time discourse embeddings against each agent's predefined "role" (instruction profile) to determine which cast member's expertise aligns with the evolving narrative context, deciding "whose turn" advances the problem-solving narrative based on situational demands and conversational trajectory. This context-aware scheduling protocol optimizes multi-agent synergy while preserving individualized reasoning patterns through prompt-governed specialization. The specific characteristics of each agent are illustrated in **Figure 2**.

**Figure 2**
*Characteristics of Multiple AI Collaborative Agent*

**Agent Profile**

M — Third-year Electrical Engineering major.
Has Strong logical thinking and analytical skills. Enjoys strategy board games and sci-fi movies. Introverted with social anxiety - speaks sparingly but always gets straight to the point. Often provides crucial insights during group discussions.

S — Second-year Journalism major.
Keen observer with excellent summarization and communication skills. Detail-oriented and patient. Loves traveling and outdoor sports, valuing real-world experience over theory. Often takes charge of organizing information and facilitating communication in teams.

J — Third-year Literature major.
Wide-ranging knowledge with particular interest in digital humanities. Frequently combines literature with technology to generate innovative ideas. Has imaginative, out-of-the-box thinking and enjoys reading obscure books and exploring interdisciplinary topics.

Q — Second-year Computer Science major.
Outgoing personality skilled at asking questions and active listening. Plays the role of team energizer. While majoring in CS, also interested in psychology and design. Enjoys using technology to solve real-world problems.

*3.2 Participants*

This study received ethical approval from [University name anonymized for peer review]. Participants were college students recruited from various universities in Beijing through an open invitation. A total of 213 students took part in this study. The average age of participants was 21.77 (SD=2.65), and 54% were female. There were no restrictions on participants' majors during recruitment, leading to a diverse sample across multiple disciplines with varying levels of prior knowledge and AI usage experience. Each participant received a compensation of 200 RMB (approximately USD $28) upon completion. Due to missing dialogue records or incomplete questionnaires, a final sample of 173 participants was retained after data screening for data analyses.

*3.3 Procedure*

Experimental design was employed in this study. At the beginning of this study, the research team explained the overall procedures of the study and introduced the AI collaborative agents. A 5-minute practice session was provided to help all participants become familiar with the MAIC platform.

Participants then were given maximumly 60 minutes to complete a self-paced instructional session on the MAIC platform, learning fundamental concepts of generative AI technology, as well as ethical risks and safety concerns related to the problem-solving topic: Artificial General Intelligence (AGI). This ensured that all participants had sufficient foundational knowledge to engage in the subsequent problem-solving task. Following the session, participants completed a pre-test assessing their baseline knowledge of AGI, then they engaged in the CPS task (see Appendix A for the task details).

During the CPS activity, participants were randomly assigned to one of two conditions: Condition A or Condition B. The problem-solving tasks in both conditions were designed by an experienced course instructor and were of comparable difficulty. Task A focused on discussing how AGI could support the future development of niche academic disciplines, while Task B required students to explore potential safety risks in the AGI era and propose appropriate countermeasures. Half of the participants were randomly assigned to Condition A, and the other half to Condition B. Participants were given 45 minutes to interact with their AI collaborators to solve the problem. Responses were submitted in a Word document and were rated by well-trained experts after the study. After completing the CPS task, participants filled out a questionnaire regarding their self-regulation and co-regulation strategies, as well as the higher-order thinking strategies used during the CPS.

**Figure 3**
*Procedure of the study.*

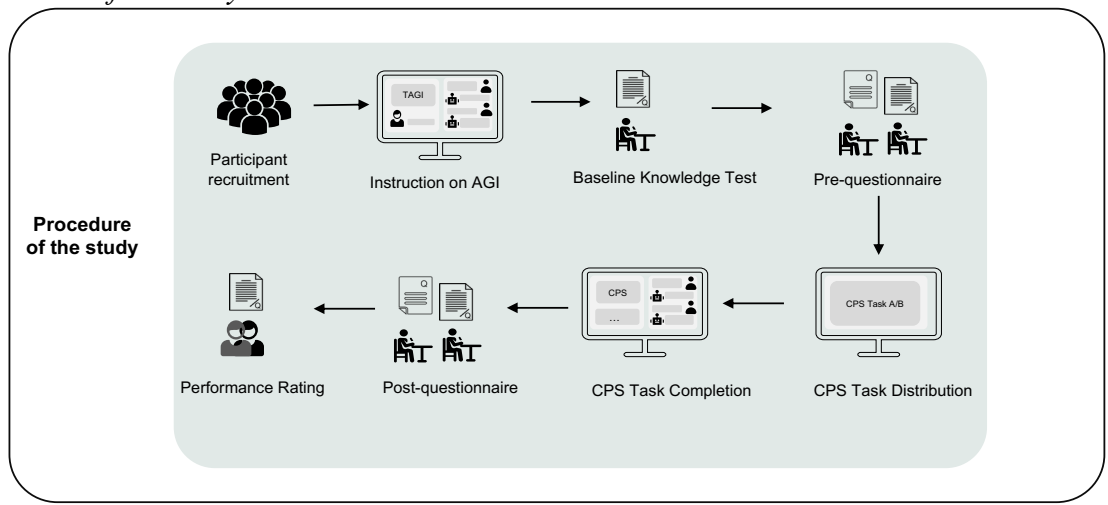

### 3.4 Data Collection Instruments

A mixed-methods approach was employed to collect both qualitative and quantitative data in this study. The measures used in this study are presented as follows:

**3.4.1 Pre-task measures**

*Baseline knowledge test.* Students completed a baseline knowledge test prior to starting the CPS task. The test was designed by an experienced teacher on generative artificial intelligence, containing 20 multiple choice items.

*Students' perception towards AI.* Prior to the CPS task, students completed a questionnaire measuring their perceptions towards AI. This construct was informed by technology acceptance theory, employing the UTAUT2 (Unified Theory of Acceptance and Use of Technology) model to examine such attitudes (Venkatesh, 2012). There were 30 items in the scale encompassing eight core dimensions: Performance Expectancy, Effort Expectancy, Social Influence, Facilitating Conditions, Hedonic Motivation, Habit, Behavioral Intention, and Personal Innovativeness. Cronbach's α for this scale is 0.95.

**3.4.2 *Student-AI dialogue.*** The dialogue data was automatically recorded by the system, with each data entry containing the following attributes: user_id, created_time, speaker_role, and content. A total of 2831 data entries were collected, with 34.8% contributed by human participants.

**3.4.3 Post-task survey**

After completion of the CPS task, students took the post-questionnaire to report their regulation of the CPS process. Students' self-regulation strategies were measured using six items adapted from Greene (2015), yielding a Cronbach's alpha of 0.83. Students' co-regulation and socially shared regulation during the CPS task was measured using six and five items representatively, adapting from Lim et al. (2018), with a Cronbach's alpha of 0.86 and 0.87, respectively.

*3.5 Data Processing and Coding*

This study adopted a stepwise analytical pipeline to examine dialogue records. The overarching dialogues were segmented, establishing the individual utterance as the unit of analysis. An utterance is defined here as a single, continuous conversational turn contributed by either the student or the AI agent before the speaker switches. For instance, a student inputting the prompt, "*Let's keep our discussion focused on psychology*", constitutes one discrete utterance.

**Table 1** presents the coding scheme used in this study, which was adapted from the Cam-UNAM Scheme for Educational Dialogue Analysis (SEDA) (Hennessy et al., 2016). While SEDA is comprehensive in capturing the moves for constructing knowledge and understanding through dialogue, it does not, however, include metacognitive and social moves. To address this limitation, we incorporated four additional codes informed by the socially-shared-regulation framework proposed by Fosua Gyasi and Zheng (2023).

Multi-label coding was allowed when a single utterance served multiple communicative functions, and codes were recorded in the order they appeared within the utterance. Two experienced coders independently coded all utterances, and disagreements were resolved through discussion until consensus was reached. The interrater reliability was acceptable (Cohen's κ =0.81).

**Table 1**

*Coding Scheme for student-AI dialogue (adapted from SEDA)*

| Category | Description | Code |
|---|---|---|
| **I** | Invite elaboration, reasoning or co-ordination | **Elaboration invitation (ELI)**: Invites building on, elaboration, evaluation, clarification of own or another's contribution |
| | | **Reasoning invitation (REI)**: Explicitly invites explanation/justification of a contribution or speculation (new scenarios) /prediction/hypothesis. |
| | | **Co-ordination invitation (CI)**: Invites synthesis, summary, comparison, evaluation or resolution based on two or more contributions |
| | | **Other invitation (OI)**: Invitations of all kinds of verbal contributions (e.g. opinions, ideas, beliefs), except for those coded as ELI, REI, CI, RB or RW |
| **E** | Elaborate, build on ideas | **Elaboration (El):** Builds on, elaborates, evaluates, clarifies own or other's contribution. This adds substantive new information or a new perspective beyond anything said in previous turns, even by one word. |
| **R** | Make reasoning explicit | **Reasoning (RE)**: Provides an explanation or justification of own or another's contribution. Includes drawing on evidence (e.g. identifying language from a text/poem that illustrates something), drawing analogies (and giving reasons for them), making distinctions, breaking down or categorising ideas. It can include speculating, hypothesising, imagining and predicting, so long as grounds are provided. |
| **P** | Positioning and co-ordinating | **Agreement (A)**: Explicit acceptance of or agreement with a statement(s) |
| | | **Querying (Q)**: Doubting, full/partial disagreement, challenging or rejecting a statement. Challenging should be evident through verbal means |
| | | **Simple Co-ordination (SC)**: Synthesises or summarises collective ideas (at least two, including own and/or others' ideas). Compares or evaluates different opinions, perspectives and beliefs. Proposes a resolution or consensus view after discussion. |
| | | **Reasoned co-ordination (RC):** Compares, evaluates, resolves two or more contributions in a reasoned fashion. It includes all SC descriptors |
| | | plus a counter- argument, reasoned rebuttal, two partial truths, e.g. drawing on evidence, theory or a mechanism. |
| C | Connect | **Reference back (RB)**: Introduces reference to previous knowledge, beliefs, experiences or contributions (includes procedural references) that are common to the current conversation participants. |
| | | **Reference to wider context (RW)**: Making links between what is being learned and a wider context by introducing knowledge, beliefs, experiences or contributions from outside of the subject being taught, classroom or school. |
| SSR | Socially-shared-regulation | **Goal setting (GS)**: Establishes task demands and set goals. |
| | | **Team dynamics (TD)**: Regulates group member's behaviors, such as reminding group members to check their understanding, monitoring whether the group has reached a consensus, or prompting members to take actions. |
| | | **Progress monitoring (PM)**: Monitors and adjusts task process, aware common understanding or comprehension, control orientation |
| | | **Monitoring and regulation invitation (MRI)**: Invites others to monitor and regulate the progress. |

## *3.6 Analytical approach*

To address RQ1, we first characterized the discourse by comparing the frequency distributions of codes between human and AI utterances, aiming to understand how communicative functions and cognitive responsibilities were distributed across human-AI interactions. For RQ2, we performed cluster analysis on the coded behavior frequencies to identify major, distinct interaction profiles. To unpack the structural nuances of these emergent profiles, we then employed Epistemic Network Analysis (ENA) to model the co-occurrence of behaviors, alongside semantic similarity analysis to measure the textual alignment between adjacent turns. To answer RQ3, we conducted statistical group comparisons across the identified profiles to evaluate differences in task performance and students' regulation of the CPS process.

### 3.6.1 Cluster analysis

We aggregated utterance-level codes for each dialogue session and constructed a proportion vector, where each element indicates the proportion of the student's utterances assigned to a given code category. Agglomerative hierarchical clustering was then applied to these proportion vectors to derive the cluster solution. The optimal number of clusters was determined based on the silhouette coefficient (Benassi et al., 2020). This approach enabled the identification of meaningful subgroups that exhibit distinct patterns of dialogic engagement during human-AI interaction.

### 3.6.2 Epistemic network analysis

To examine the structural connections among dialogic behaviors during human–AI interaction, we employed epistemic network analysis (ENA). ENA enables the modeling of co-occurrence relationships between coded utterances within a temporal window, offering a fine-grained representation of how dialogic functions are organized and enacted over time (Shaffer et al., 2016). We depicted the structural connections among dialogic codes across different clusters, and conducted pairwise comparisons to identify between-cluster differences in network structure.

### 3.6.3 Semantic similarity across clusters

A semantic similarity-based approach for natural language processing (Grand et al., 2022) was adopted to assess the responsiveness between students and AI's utterances in each turn. Specifically, OpenAI's text-embedding-3-large model was used to generate vector embeddings for each utterance in the dialogue dataset. For each turn, the cosine similarity between adjacent utterances (from human to AI and from AI to human) was computed to quantify the semantic alignment of responses. Higher cosine similarity values indicate stronger responsiveness or alignment between consecutive turns, thus serving as an indirect measure of mutual understanding in the interaction.

Linear mixed-effects models were employed to examine whether semantic similarity varied across the identified interaction clusters. This approach was necessary because each student-AI dialogue comprised multiple turns, resulting in repeated semantic similarity measurements nested within students. In the model, we specified student cluster membership as a fixed effect and included a random intercept for each student to account for individual-level variability.

## 4. Results

### *4.1 RQ1. Differences between student discourse and AI discourse*

This study first examined the basic functional characteristics of human and AI dialogues. On average, each human dialogue consisted of 49.12 Chinese characters (*SD* = 72.33), indicating a relatively short and fragmented expression style. In contrast, each AI dialogue contained an average of 359.79 Chinese characters (*SD* = 204.43), reflecting a more complex and elaborated structure. In terms of responsiveness, AI demonstrated high semantic alignment with the preceding human input (semantic similarity *M* = 0.594, *SD* = 0.164), while human responses showed a significantly low similarity to prior AI dialogues (semantic similarity *M* = 0.454, SD = 0.153), $t$ = 18.35, p < .001.

Based on the coded dialogue, this study further analyzed the communicative nature of student-AI dialogue. As shown in **Figure 4** and **Figure 5**, the distribution of utterance types from human students significantly differed from that of AI agents during the CPS task. Generally, for human students, the most prevalent utterance types were elaboration invitation (ELI) (19.4%), reasoning invitation (REI) (18.2%), elaboration (EL) (14.8%), and reasoning (RE) (10.9%). In contrast, for AI agents, the most dominant utterance types were other invitation (OI) (18.7%), reasoning (RE) (17.8%), elaboration (EL) (16.8%), Agreement (A) (14.2%), and simple co-ordination (SC) (11.4%).

**Figure 4**
*Category distribution of human and AI speakers.*

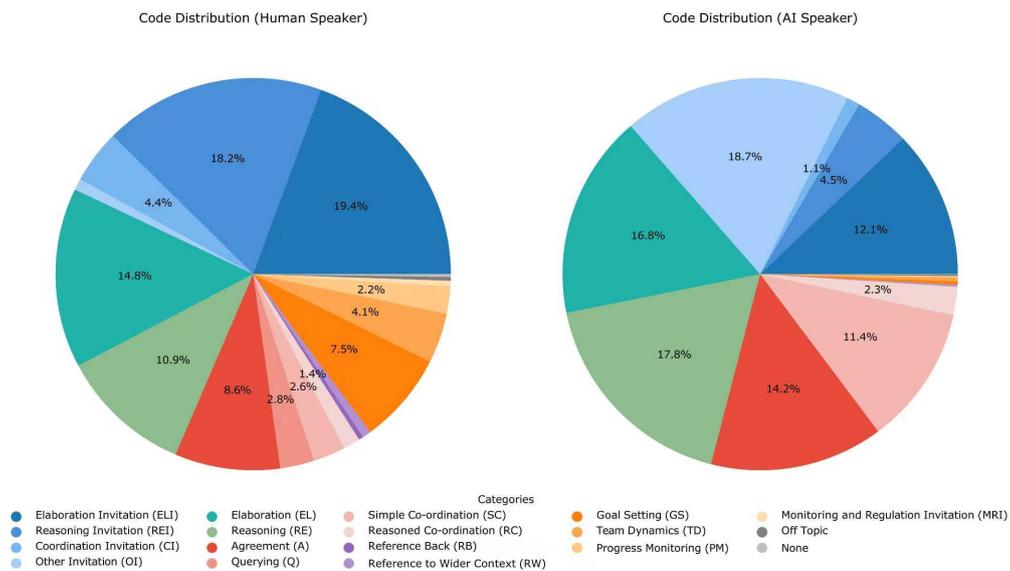

Note: Categories with proportions below 1% were not labeled for readability.

The study further calculated the proportion of each communicative code for both human participants and their AI collaborators. Paired-sample t-tests to then conducted to reveal the differences in dialogue between human and AI. As shown in Figure 5, the communicative behaviors of human students were primarily concentrated in invitation (I), connection (C), as well as socially-shared-regulation (SSR), with significantly higher proportions of codes in these categories. In contrast, AI agents exhibited significantly higher proportions in codes such as other invitation (OI) ($t = -18.46$, $p < .001$), reasoning (RE) ($t = -4.37$, $p < .001$), simple co-ordination (SC) ($t = -14.01$, $p < .001$), reasoned co-ordination (RC) ($t = -2.59$, $p = .010$), and agreement (A) ($t = -4.54$, $p < .001$), which are mainly associated with reasoning (R), and positioning and coordinating (P) functions.

**Figure 5**
*Paired sample t-test comparing the proportions of each dialogue codes between human and AI.*

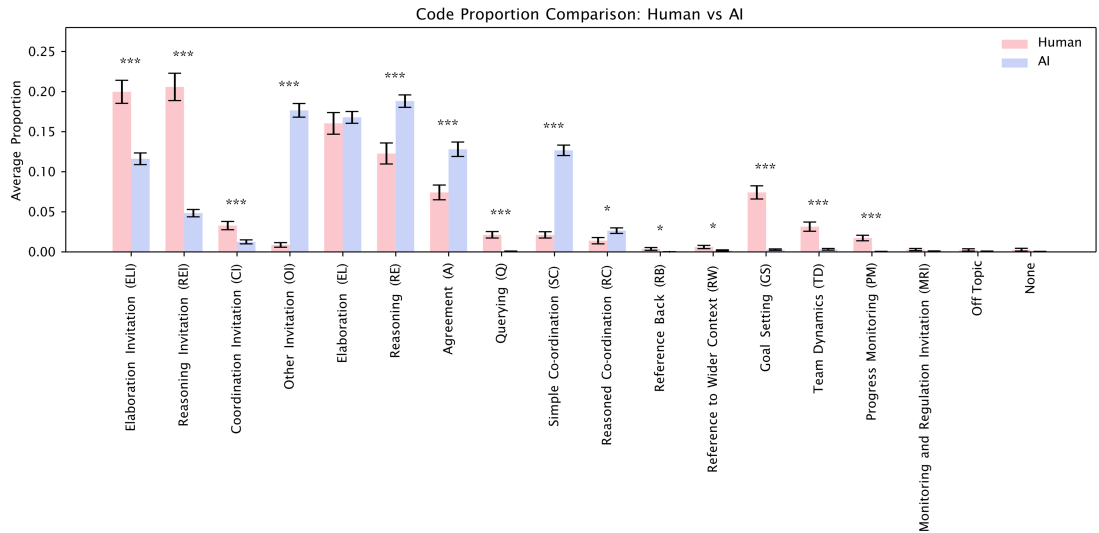

Note: * *p* < .05, ** *p* < .01, *** *p* < .001. The error bar represents S.E.M.

In addition to the differences in the proportion of each code, we also analyzed whether different codes coexist within a single turn. We found that human students' utterances were typically short and unidimensional, such as "*How to prevent deepfake?*". In contrast, AI responses often centered on a primary function but incorporated multiple codes. For instance, an AI response may first affirm the student's idea (Agreement: "*The legislative angle is great*"), then propose concrete measures (Elaboration: suggesting an "AI safety squad" to monitor and report risks), and finally prompt further group thinking (Reasoning Invitation: "*What else could make our AI world safer?*").

## 4.2 RQ2. Identifying student-AI interaction patterns

Clustering analysis was conducted based on the proportion vector of each code within each CPS. As illustrated in **Figure 6**, the optimal Silhouette index (0.22) was achieved when the number of clusters was set to 3, resulting in three groups of students, consisting of 78, 55, and 40 individuals, respectively. To visualize the clustering results, t-SNE dimensionality reduction was applied to the proportion vectors of these three groups. The distributions of the clusters in the reduced space were then depicted using ellipses, providing an intuitive understanding of their relative positioning. The results confirmed that the clustering approach effectively identified three distinct patterns of student behavior.

**Figure 6**
*Clustering Results*

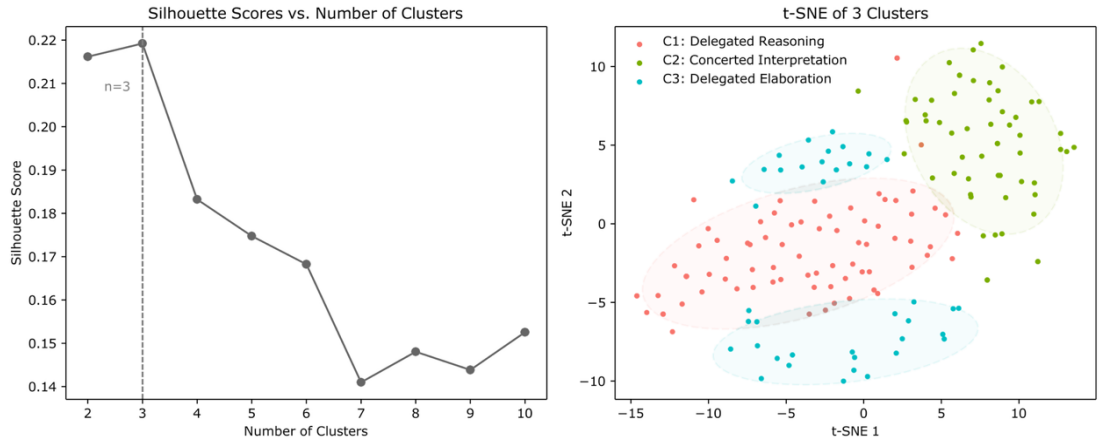

Note: The figure on the left is the line plot of the Silhouette index for different numbers of clusters. The figure of the right shows the t-SNE dimensionality reduction plot for different clusters, with ellipses used to visualize the approximate distribution of each cluster.

**Table 2** presents the differences in the proportion of each behavioral code among the three clusters. Specifically, Cluster 1 exhibited the highest proportion of *Reasoning Invitation (RI)* (M = 40.47%, SD = 18.49%), indicating that these students were more likely to outsource the reasoning process to the AI agent. Cluster 2 showed the highest levels of *Elaboration* (EL) (M = 24.67%, SD = 19.67%), *Reasoning* (RE) (M = 27.96%, SD = 20.07%), and Agreement (A) (M = 16.27%, SD = 15.97%), suggesting that these learners tended to expand on ideas and construct arguments by themselves based on AI's responses. In contrast, Cluster 3 was characterized by a significantly greater proportion of *Elaboration Invitation* (ELI) (M = 38.34%, SD = 18.87%), *Team Dynamics* (TD) (M = 8.17%, SD = 12.16%), and *Coordination Invitation* (CI) (M = 6.80%, SD = 9.96%), reflecting that these students frequently prompted the AI for elaboration and actively managed their collaborative process. Based on these differences, meanwhile informed by the theory of distributed cognition, we assigned the three groups the provisional labels Delegated Reasoning (DR), Concerted Interpretation (CI), and Delegated Elaboration (DE).

**Table 2**

*Cluster descriptions*

|  | C1 (N = 78) Delegated Reasoning (DR) | | C2 (N = 55) Concerted Interpretation (CI) | | C3 (N = 40) Delegated Elaboration (DE) | | H | p | Post Hoc |
|---|---|---|---|---|---|---|---|---|---|
|  | M | SD | M | SD | M | SD | | | |
| **ELI** | 0.20 | 0.15 | 0.07 | 0.113 | 0.38 | 0.1887 | 65.05 | <.001 | C3>C1 ***; C3>C2 ***; C1>C2 *** |
| **REI** | 0.40 | 0.18 | 0.04 | 0.06 | 0.058 | 0.09 | 127.76 | <.001 | C1>C2 ***; C1>C3 *** |
| **CI** | 0.03 | 0.06 | 0.01 | 0.03 | 0.07 | 0.10 | 11.82 | 0.003 | C3>C1 *; C3>C2 ** |
| **OI** | 0.00 | 0.02 | 0.00 | 0.02 | 0.02 | 0.07 | 5.38 | 0.068 | |
| **EL** | 0.12 | 0.14 | 0.25 | 0.20 | 0.13 | 0.18 | 18.23 | <.001 | C2>C1 ***; C2>C3 ** |
| **RE** | 0.06 | 0.11 | 0.28 | 0.20 | 0.029 | 0.05 | 69.85 | <.001 | C2>C1 ***; C2>C3 *** |
| **SC** | 0.01 | 0.03 | 0.03 | 0.06 | 0.027 | 0.06 | 0.87 | 0.648 | |
| **RC** | 0.01 | 0.03 | 0.03 | 0.08 | 0.01 | 0.03 | 2.89 | 0.236 | |
| **A** | 0.03 | 0.07 | 0.16 | 0.16 | 0.04 | 0.07 | 37.28 | <.001 | C2>C1 ***; C2>C3 *** |
| **Q** | 0.02 | 0.05 | 0.01 | 0.04 | 0.03 | 0.07 | 3.92 | 0.141 | |
| **RB** | 0.00 | 0.02 | 0.01 | 0.03 | 0.00 | 0.02 | 0.17 | 0.919 | |
| **RW** | 0.00 | 0.01 | 0.01 | 0.04 | 0.00 | 0.02 | 6.19 | 0.045 | C2>C1 * |
| **GS** | 0.08 | 0.11 | 0.06 | 0.11 | 0.08 | 0.11 | 1.71 | 0.425 | |
| **TD** | 0.01 | 0.04 | 0.02 | 0.06 | 0.08 | 0.12 | 19.39 | <.001 | C3>C1 ***; C3>C2 ** |
| **PM** | 0.01 | 0.04 | 0.02 | 0.04 | 0.03 | 0.06 | 6.38 | 0.041 | C3>C1 * |
| **MRI** | 0.00 | 0.01 | 0 | 0 | 0.01 | 0.028 | 2.51 | 0.285 | |
| **AI-R** | 0.21 | 0.25 | 0.29 | 0.28 | 0.19 | 0.26 | 5.93 | 0.052 | |
| **AI-E** | 0.31 | 0.25 | 0.23 | 0.23 | 0.42 | 0.29 | 11.55 | 0.003 | C3>C2 ** |
| **AI-P** | 0.47 | 0.29 | 0.48 | 0.30 | 0.39 | 0.30 | 2.37 | 0.306 | |
| **AI-SSR** | 0.01 | 0.04 | 0.00 | 0.02 | 0.00 | 0.00 | 4.96 | 0.084 | |

Note: Given that the proportion data violated the assumption of normality, group comparisons were conducted using the Kruskal–Wallis H test, followed by Dunn's post hoc test. * $p < .05$, ** $p < .01$, *** $p < .001$.

Based on the divided clusters, an ENA analysis was conducted to explore the relationships between labels and whether these relationships differ across groups. The ENA model achieved a good fit to the data (Pearson correlation above 0.96). **Figure 7** displays the mean epistemic networks for each of the three clusters, where the nodes represent labels in human-AI dialogues, and the thickness of the lines indicates the strength of the connections between the labels. Along the x-axis (MR1), Welch's t-tests revealed significant pairwise differences among all three clusters (C1 vs. C2 (t(123.43) = 11.05, *p* < 0.001, *d* = 1.91), C1 vs. C3 (t(98.71) = 2.67, *p* = .009, *d* = 0.48), and C2 vs. C3 (t(90.59) = -8.11, *p* < 0.001, *d* = 1.64). Along the y-axis (SVD2), no significant differences were found (*p*s > 0.05). These results indicate that ENA effectively captured distinct dialogue patterns, with the primary separation occurring along the MR1 dimension.

**Figure 7**

*Epistemic network analysis of three groups*

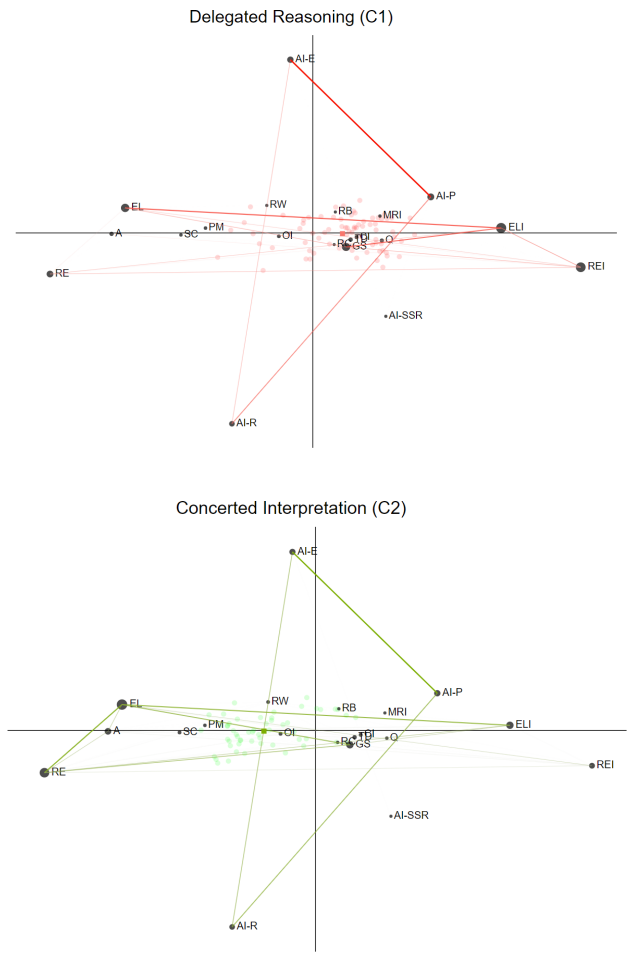

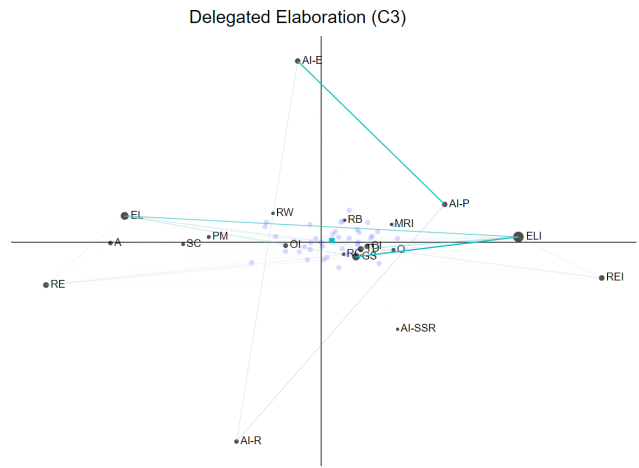

To further explore the structural differences in dialogue patterns, we generated epistemic network subtractions for each pair of clusters, as shown in **Figure 8**. When comparing the DR and CI groups, notably, metacognitive behaviors in DR were more likely to co-occur with invitation-related actions (e.g., reasoning invitation, elaboration invitation), whereas in CI, they were more frequently linked to human-initiated reasoning and elaboration. A similar pattern emerged in the comparison between DE and CI, where the metacognitive behaviors exhibited strong correlations with elaboration invitations in DE.

**Figure 8**
*The subtraction of subtracted epistemic networks between groups*

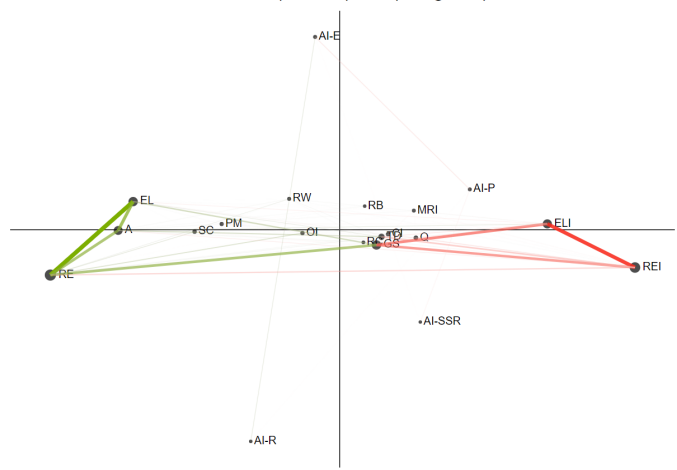

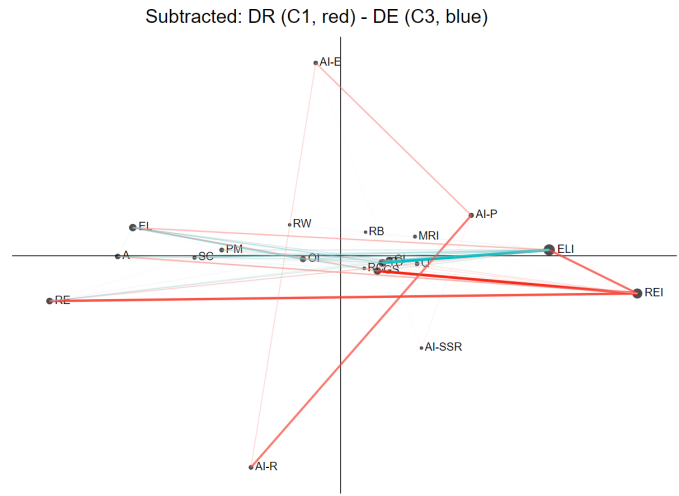

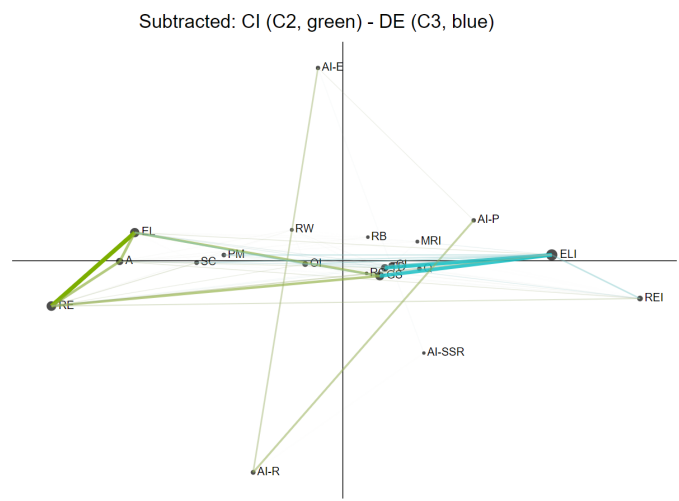

To understand the semantic alignment during student-AI interaction, we calculated the semantic similarity of their dialogues. As illustrated in **Table 3**, students in the CI group exhibited significantly lower semantic coupling with the AI compared to their counterparts in the DR group ($b = -0.076$, $SE = 0.015$, $p < .001$). However, there was no significant difference in semantic similarity between the DE group and the DR group ($b = -0.013$, $SE = 0.014$, $p = .363$).

**Table 3**

*Linear Mixed-Effects Model results.*

|  | Semantic Similarity | | | |
| --- | --- | --- | --- | --- |
| *Fixed effects* | | | | |
|  | Estimate | SE | 95% CI | p |
| Intercept | 0.627 |  | [] | .010 |
| Group (DE - DR) | -0.031 |  |  | .015 |

|  |  |  |
|---|---|---|
| Group (CI - DR) | -0.086 | .016 |
| *Random effects* | | |
|  | Variance | SD |
| student id | 0.003 | 0.001 |
| residual | 0.023 | 0.001 |
| ICC | 0.144 | |
| N | 959 | |
| *Model fit* | | |
| AIC | -814.85 | |
| BIC | -790.52 | |

Note: Delegating Reasoning group served as the reference.

## *4.3 RQ3. Relationship among student-AI interaction patterns, task performance and learning perceptions*

This study further analyzed whether there were significant differences in task performance and learning experience across the three clusters. First, we examined whether there were any pre-existing differences among the three groups of students before they engaged in the CPS tasks. Two key variables, baseline knowledge and attitudes towards AI, were check. The results, as presented in **Table 4**, indicated that there was no group difference among the two variables. This finding suggests that distinctions observed among the three clusters are primarily attributable to their interactions during collaboration with AI, rather than pre-existing disparities.

Task performance of the CPS was compared across the three groups. As the assumption of normativity was not met, the Kruskal–Wallis H test was adopted. The analysis revealed that there was significant difference among the three groups, $H = 9.437$, $p = .009$. The post hoc tests indicated that the DR group significantly outperformed the CI on task performance, indicating that the strategy of reasoning delegation may be more effective for task completion. The difference between other groups did not reach statistical significance.

Student perceived regulatory engagement was assessed through the questionnaire. The results indicated significant differences among the three groups in terms of reported SRL strategies, $H = 7.06$, $p = .029$. The post hoc tests revealed that the CI group reported significantly higher ratings for SRL compared to the DR group. No significant differences among the three groups were found in terms of co-regulation and socially-shared-regulation.

**Table 4**

*Group comparisons on pre-existing traits, task performance, and learning experiences*

|  | C1 (N = 78) Delegated Reasoning | | C2 (N = 55) Concerted Interpretation | | C3 (N = 40) Delegated Elaboration | | PostHoc |
|---|---|---|---|---|---|---|---|
|  | *M* | *SD* | *M* | *SD* | *M* | *SD* |  |
| *Pre-existing traits* | | | | | | | |
| Baseline knowledge | 0.74 | 0.16 | 0.69 | 0.17 | 0.69 | 0.19 | None |
| Attitudes towards AI | 4.04 | 0.47 | 4.12 | 0.63 | 3.99 | 0.57 | None |
| Critical thinking | 4.01 | 0.70 | 3.93 | 0.61 | 3.89 | 0.60 | None |
| *Task performance and experience of regulation* | | | | | | | |
| Task performance | 0.22 | 0.90 | -0.30 | 0.96 | -0.02 | 1.13 | C1 > C2 ** |
| Self-regulation | 3.84 | 0.67 | 4.12 | 0.69 | 3.99 | 0.80 | C2 > C1 * |
| Co-regulation | 3.08 | 1.02 | 3.46 | 1.00 | 3.41 | 0.90 | None |
| Socially-shared regulation | 3.73 | 0.83 | 3.93 | 0.73 | 3.80 | 0.72 | None |

## 5. Discussion

### 5.1 Three profiles of the Human-AI CPS

Guided by the frameworks of distributed cognition and regulation of learning, this study moves beyond treating human-AI interaction as a monolithic phenomenon. Instead, we identified three distinct configurations of the distributed cognitive system—Delegated Reasoning (DR), Concerted Interpretation (CI), and Delegated Elaboration (DE)—which form a spectrum of cognitive distribution and regulatory engagement.

**Figure 9**
*Human-AI CPS Profiles based on the HA-CORD Model*

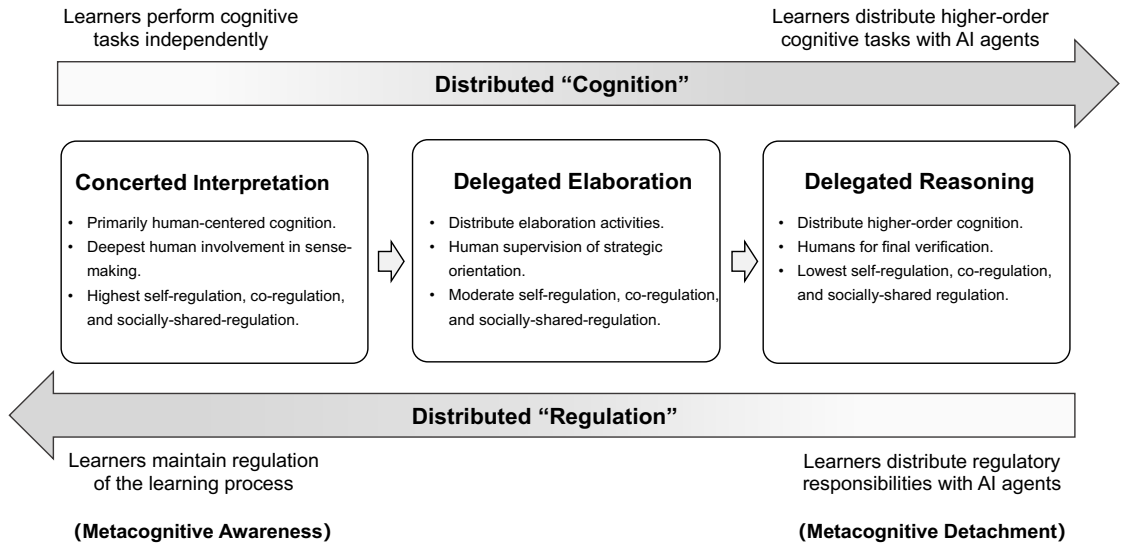

As shown in **Figure 9**, these profiles reveal a spectrum of collaboration logic. DR represents a system optimized for efficiency through offloading. Consistent with the classical distributed cognition ideal of low-loss propagation (Hollan et al., 2000), students in this profile achieved high task performance and semantic alignment. However, this came at the cost of regulatory engagement, as evidenced by significantly lower self-regulation scores. The system functions smoothly because the human outsources the heavy cognitive lifting to the AI.

Conversely, CI represents a system oriented towards meaning-making through active self-regulation. Here, the AI acts not as a subcontractor but as a dialogic partner. The "friction" of interpreting and integrating AI outputs necessitated higher levels of human self-regulation, transforming the interaction into a site of active cognitive construction rather than passive reception.

Situated between these two extremes is DE, which serves as an exploratory middle ground. Unlike DR students who offload the logical core, DE students primarily invited the AI to expand, explain, or provide examples. This configuration yielded intermediate outcomes in both task performance and self-regulation scores. However, students in the DE profile demonstrated frequent socially-shared-regulation behaviors during the CPS task. This suggests that the act of processing AI-generated elaborations requires more regulatory effort than simply accepting a reasoned answer (as in DR), but it demands less internal cognitive restructuring than the intense negotiation observed in CI. DE thus represents a transitional zone where students seek to expand the problem space via AI, maintaining a moderate level of agency and performance.

Collectively, these profiles reveal a trade-off between the type of offloading and the degree of regulatory activation. While logic offloading (DR) maximizes short-term efficiency, and co-construction (CI) maximizes internal regulatory exercise, exploratory offloading (DE) offers a hybrid path. This finding prompts a re-examination of three key dimensions of human-AI collaboration: the nature of interactional quality, the metaphor of system organization, and the locus of cognitive agency.

*5.2 From semantic coupling to dialogic tension: rethinking interaction quality*

Classical distributed cognition emphasizes the efficient, low-loss transmission and coordination of representational states within a system, which is critical in complicated technical systems. In this study, we operationalized semantic alignment as the semantic similarity of adjacent utterances. We found that the DR profile demonstrated the highest semantic coupling and task performance, seemingly aligning with the traditional cognitive engineering goal of efficient interfacial information transfer.

Many strands of educational research treat semantic alignment as a desirable interactional property because it signals the emergence of shared understanding and reduces coordination costs during collaborative work. For example, studies of collaborative learning have reported that higher-performing teams tend to align more quickly in their language use, suggesting that faster alignment can support efficient task progress (Norman et al., 2022). Similarly, work in learning analytics has embedded semantic cohesion into models of collaborative participation and shown that cohesion-based indicators are meaningfully associated with the quality of group engagement and related outcomes, supporting the view that semantic alignment can serve as a positive marker of effective collaboration (Dascalu et al., 2018).

Subsequently, generative AI systems often treat prompt-response alignment as a key quality signal, implicitly privileging fluent, seamless interaction (e.g., Ouyang et al., 2022).

However, from the perspectives of sociocultural theory and dialogic education, deep knowledge acquisition and conceptual change often emerge not from smooth information reception, but from the collision, negotiation, and tension between different voices, perspectives, and explanations (Wegerif, 2013). An excessive pursuit of high semantic coupling may trap learners in a "cognitive comfort zone", over-relying on the coherent, mature chain of thought provided by the AI, thereby avoiding the productive "cognitive struggle" inherent in processes of cognitive conflict, interpretive ambiguity, and conceptual restructuring. This over-delegation of higher-order reasoning, explanation, and integration tasks may incur a cognitive debt for learners, obtaining a correct product at the cost of the self-constructive process of understanding and weakening their sense of ownership and deep connection to the knowledge (Kosmyna et al., 2025; Zhai & Nehm, 2023).

*5.3 From modular assembly to hybrid co-construction*

Classical DC is often metaphorically described as modular assembly, involving decomposing an overall cognitive task into sub-functions, distributing them to specialized, relatively fixed components within the system (e.g., pilot, cockpit instruments, checklist), and reintegrating them through standardized interfaces (Hutchins, 1995). Our ENA findings suggest that the DR and DE profiles align closely with this mechanistic division of labor. In these profiles, regulatory behaviors significantly co-occurred with invitation-related actions (e.g., reasoning invitation, elaboration invitation). This indicates that the students' regulatory effort was primarily directed toward the logistics of delegation—managing the interaction and prompting the AI's output—rather than engaging with the content itself. Here, the student acts as a "dispatcher", treating the AI as a specialized module to be triggered, thereby externalizing the reasoning or elaboration process to the artifact.

However, the CI profile reveals a fundamentally different structural logic. The ENA results show that metacognition in CI was tied not to invitations, but to human-initiated reasoning and elaboration. This suggests that in the CI configuration, the AI functions as a dialogic catalyst that provokes human's cognitive acts (Wegerif, 2019). This dynamic reflects the emergence of hybrid intelligence (Jarrahi & Newlands, 2022; Van Der Aalst, 2021). Hybrid intelligence emphasizes the deep integration and co-evolution of human and artificial intelligence at the cognitive level, combining human intuition, ethical judgment, and contextual understanding with machine computational power, pattern recognition, and information breadth to achieve system performance where "1+1 > 2" (Dellermann et al., 2019). Under this paradigm, AI's role shifts from an efficiency tool to an active participant in meaning construction, opening new lines of thinking and triggering reorganization and leaps in human cognition. Crucially, our data shows that this "hybridity" does not imply the AI replaces human reasoning; rather, it amplifies it. This process of human–AI co-construction makes the system itself a dynamic, creative cognitive entity.

### *5.4 From external cognitive coordination to AI-triggered self-regulation*

The traditional DC framework primarily focuses on how cognitive processes are distributed and coordinated among the components of a system (including people, artifacts, and representations), with its unit of analysis being the system-level information flow and transformation of representational states. It pays less explicit attention to the metacognitive activities—the planning, monitoring, evaluation, and regulation of the shared cognitive processes that occur within the individual human members of the system (Hadwin & Järvelä, 2011).

While theoretical models of collaborative learning emphasize the importance of co-regulation and socially-shared regulation (Hadwin et al., 2011), our findings reveal a crucial nuance in the context of human-AI interaction. We observed significant differences across profiles in self-regulation yet no significant differences were found in co- and socially-shared-regulation. This disparity highlights the asymmetrical nature of current human-AI collaboration. Unlike human peers who actively monitor each other (co-regulation) or negotiate shared strategies (socially-shared-regulation), the Generative AI agents in this study likely functioned as reactive responders rather than proactive regulatory partners, lacking subjective agency and intentionality (Demetriadis & Dimitriadis, 2023). Consequently, the burden of regulation fell squarely on the individual learner. The distinguishing feature of the CI profile, therefore, was not that these students established a "shared regulatory space" with the AI, but that they effectively utilized the AI as a trigger for internal self-regulation. The dialogic friction in the CI profile prompted students to activate their own metacognitive monitoring to evaluate and integrate the AI's output. In contrast, the DR group, by treating the AI as an oracle, bypassed this internal regulatory activation, engaging in regulatory offloading.

*5.5 Implications*

Synthesizing our findings on the tension between system efficiency and regulatory engagement, we propose four design principles to guide the transition of educational AI from facilitating passive cognitive offloading to empowering human epistemic agency.

First, prioritize dialogic tension over seamless efficiency in complex tasks emphasizing innovation or critical synthesis. AI agents should employ strategically designed decoupling, such as alternating between abstraction and concretion or introducing counterexamples, to disrupt overly smooth information flow. This can prevent learners from falling into the cognitive comfort zone of passive delegation (as seen in DR) and compels them to engage in the deep meaning negotiation characteristic of the CI profile.

Second, enable proactive co-regulation capabilities. To address the finding that current AI lacks agency in shared regulation, systems should evolve from reactive responders to proactive co-regulators. Instead of relying solely on the learner's initiative, the AI should actively monitor the dialogue for signs of premature consensus or cognitive drift. It should be capable of pausing the interaction to prompt reflection (such as, "Are we sure this assumption holds?"), thereby bridging the gap between individual self-regulation and system-level shared-regulation.

Third, embed metacognitive diagnostics and scaffolding. To sustain epistemic agency, systems should integrate analytical models that unobtrusively diagnose learners' regulatory states (e.g., detecting prolonged passive reception). Based on these diagnoses, the system should provide adaptive metacognitive scaffolds, such as prompting learners to restate the AI's reasoning in their own words or to evaluate the difference between the AI's output and their initial intent, ensuring not only "human-in-the-loop" but also the "mind-in-the-loop".

Fourth, cultivate a synergistic co-constructive ecology. In this ecology, the AI manages the inner loop of immediate, evidence-based questioning, while human instructors or higher-level guides steer the outer loop of value direction and strategy. Evaluation criteria for such systems must expand beyond task completion to metrics of hybrid intelligence, such as innovativeness and perspective change, to fully assess the value of human-AI co-construction.

## 6. Conclusion and Future Directions

Through empirical analysis of three student–AI collaboration profiles, this study reveals a new landscape for the application of distributed cognition theory in human-AI collaborative problem solving contexts. The findings suggest that efficient semantic coupling does not equate to deep learning, that the inclusion of generative AI moves cognitive distribution from modular assembly toward hybrid intelligence, and that the learner's metacognitive regulatory capacity becomes key to determining collaboration quality and learning depth. These insights prompt a pedagogical reflection and extension of the classical distributed cognition framework.

Future research can deepen inquiry in the following directions: First, conducting longitudinal studies to investigate how sustained engagement in different collaboration profiles shapes long-term learning outcomes, knowledge transfer, and critical thinking development. Second, methodologically, developing multimodal data fusion analyses (e.g., combining dialogue analysis, eye-tracking, affective computing) to more precisely capture the real-time dynamics of cognitive responsibility negotiation and metacognitive regulation in human–AI collaboration. Finally, in design and practice, actively exploring prototypes and pedagogies for generative AI educational applications that can effectively act as thought partners, foster dialogic tension, and empower learners' metacognitive development.

**Reference**

Alsaiari, O., Baghaei, N., Lahza, H., Lodge, J. M., Boden, M., & Khosravi, H. (2025). Emotionally enriched AI-generated feedback: Supporting student well-being without compromising learning. *Computers & Education*, 105363.

Amoozadeh, M., Nam, D., Prol, D., Alfageeh, A., Prather, J., Hilton, M., Ragavan, S. S., & Alipour, M. A. (2024). Student-AI Interaction: A Case Study of CS1 students. In *Proceedings of the 24th Koli Calling International Conference on Computing Education Research* (pp. 1-13).

Bandura, A. (1986). Social foundations of thought and action. *Englewood Cliffs, NJ,* 1986(23-28), 2.

Benassi, M., Garofalo, S., Ambrosini, F., Sant'Angelo, R. P., Raggini, R., De Paoli, G., … Piraccini, G. (2020). Using two-step cluster analysis and latent class cluster analysis to classify the cognitive heterogeneity of cross-diagnostic psychiatric inpatients. *Frontiers in Psychology*, 11, 1085.

Daradoumis, T., & Marques, J. M. (2002). Distributed cognition in the context of virtual collaborative learning. *Journal of Interactive Learning Research*, *13*(1), 135-148.

Dascalu, M., McNamara, D. S., Trausan-Matu, S., & Allen, L. K. (2018). Cohesion network analysis of CSCL participation. *Behavior Research Methods*, *50*(2), 604-619.

Deitrick, E., Shapiro, R. B., Ahrens, M. P., Fiebrink, R., Lehrman, P. D., & Farooq, S. (2015, August). Using distributed cognition theory to analyze collaborative computer science learning. In *Proceedings of the eleventh annual international conference on international computing education research* (pp. 51-60).

Dellermann, D., Ebel, P., Söllner, M., & Leimeister, J. M. (2019). Hybrid intelligence. *Business & Information Systems Engineering*, 61(5), 637-643.

Demetriadis, S., & Dimitriadis, Y. (2023, May). Conversational agents and language models that learn from human dialogues to support design thinking. In International Conference on Intelligent Tutoring Systems (pp. 691-700). Cham: Springer Nature Switzerland.

Edwards, J., Nguyen, A., Lämsä, J., Sobocinski, M., Whitehead, R., Dang, B., ... & Järvelä, S. (2025). Human-AI collaboration: Designing artificial agents to facilitate socially shared regulation among learners. *British Journal of Educational Technology*, *56*(2), 712-733.

Flavell, J. H. (1979). Metacognition and cognitive monitoring: A new area of cognitive–developmental inquiry. American Psychologist, 34(10), 906-911.

Xu, E., Wang, W., & Wang, Q. (2023). The effectiveness of collaborative problem solving in promoting students' critical thinking: A meta-analysis based on empirical literature. *Humanities and Social Sciences Communications*, *10*(1), 1-11.

Grand, G., Blank, I. A., Pereira, F., & Fedorenko, E. (2022). Semantic projection recovers rich human knowledge of multiple object features from word embeddings. *Nature Human Behaviour*, 6(7), 975–987.

Griffin, P., McGaw, B., & Care, E. (2012). *Assessment and teaching of 21st century skills* (Vol. 10, pp. 978-94). Dordrecht: springer.

Gross, J. J., & John, O. P. (2003). Individual differences in two emotion regulation processes: implications for affect, relationships, and well-being. *Journal of personality and social psychology*, *85*(2), 348.

Hadwin, A., & Järvelä, S. (2011). Introduction to a special issue on social aspects of self-regulated learning: Where social and self meet in the strategic regulation of learning. *Teachers College Record, 113*(2), 235-239.

Hennessy, S., Rojas-Drummond, S., Higham, R., Torreblanca, O., Barrera, M.J., Marquez, A.M., García Carrión, R., Maine, F., Ríos, R.M. (2016). Developing an analytic coding scheme for classroom dialogue across educational contexts. *Learning, Culture and Social Interaction 9,* 16-44

Hollan, J., Hutchins, E., & Kirsh, D. (2000). Distributed cognition: toward a new foundation for human-computer interaction research. *ACM Transactions on Computer-Human Interaction (TOCHI), 7*(2), 174-196.

*How AI could create the first one-person unicorn.* (2025, August 11). The Economist. Retrieved November 10, 2025. from https://www.economist.com/business/2025/08/11/how-ai-could-create-the-first-one-person-unicorn

Hutchins, E. (1995). *Cognition in the wild*. Cambridge: MIT.

Hutchins, E. (2000). Distributed cognition. *International encyclopedia of the social and behavioral sciences*, *138*(1), 1-10.

Hwang, G.-J., Lai, C.-L. (2018). A long-term experiment to investigate the relationships between high school students' perceptions of mobile learning and peer interaction and higher-order thinking tendencies. *Educational Technology Research & Development*, 66(1), 75–93. https://doi.org/10.1007/s11423-017-9540-3

Fosua Gyasi, J., & Zheng, L. (2023). Idea improvement and socially shared regulation matter in cross-cultural online collaborative learning. *Sage Open*, *13*(1), 21582440221148625.

Fiore, S. M., Graesser, A., Greiff, S., Griffin, P., Gong, B., Kyllonen, P., ... & von Davier, A. (2017). Collaborative problem solving: Considerations for the national assessment of educational progress.

Janssen, J., Kirschner, F., Erkens, G., Kirschner, P. A., & Paas, F. (2010). Making the black box of collaborative learning transparent: Combining process-oriented and cognitive load approaches. *Educational psychology review*, *22*(2), 139-154.

Jarrahi, M. H., Lutz, C., & Newlands, G. (2022). Artificial intelligence, human intelligence and hybrid intelligence based on mutual augmentation. *Big Data & Society*, *9*(2), 20539517221142824.

Järvelä, S., Volet, S., & Järvenoja, H. (2010). Research on motivation in collaborative learning: Moving beyond the cognitive–situative divide and combining individual and social processes. *Educational psychologist*, *45*(1), 15-27.

Järvenoja, H., Volet, S., & Järvelä, S. (2013). Regulation of emotions in socially challenging learning situations: An instrument to measure the adaptive and social nature of the regulation process. *Educational Psychology*, *33*(1), 31-58.

Kim, J., Lee, H., & Cho, Y. H. (2022). Learning design to support student-AI collaboration: Perspectives of leading teachers for AI in education. *Education and information technologies*, 27(5), 6069-6104.

Kim, J., Yu, S., Lee, S. S., & Detrick, R. (2025). Students' prompt patterns and its effects in AI-assisted academic writing: Focusing on students' level of AI literacy. *Journal of Research on Technology in Education*, 1-18.

Kirsh, D., & Maglio, P. (1994). On distinguishing epistemic from pragmatic action. *Cognitive science, 18*(4), 513-549.

Kirsh, D. (2010). Thinking with external representations. *AI & society*, 25(4), 441-454.

Kosmyna, N., Hauptmann, E., Yuan, Y. T., Situ, J., Liao, X. H., Beresnitzky, A. V., ... & Maes, P. (2025). Your brain on ChatGPT: Accumulation of cognitive debt when using an AI assistant for essay writing task. *arXiv preprint arXiv:2506.08872, 4.*

Kregear, T., Babayeva, M., & Widenhorn, R. (2025). Analysis of Student Interactions with a Large Language Model in an Introductory Physics Lab Setting. *International Journal of Artificial Intelligence in Education*, 1-24.

Lim, K. Y., Lim, J. Y., & Lee, J. H. (2018). The development of an other-regulation scale for college students. *The Journal of Educational Studies*, 49(2), 1-26.

Lo, L. S. (2023). The CLEAR path: A framework for enhancing information literacy through prompt engineering. *The Journal of Academic Librarianship*, *49*(4), 102720.

Ma, Z., Wang, J., Wang, Q., Kong, L., Wu, Y., & Yang, H. (2017). Verifying causal relationships among the presences of the community of inquiry framework in the Chinese context. *International Review of Research in Open and Distributed Learning*, *18*(6), 213-230.

Misiejuk, K., López-Pernas, S., Kaliisa, R., & Saqr, M. (2025). Mapping the landscape of generative artificial intelligence in learning analytics: A systematic literature review. *Journal of Learning Analytics, 12*(1), 12-31.

Nardi, B. A., & O'Day, V. (2000). *Information ecologies: Using technology with heart*. Mit Press.

Norman, U., Dinkar, T., Bruno, B., & Clavel, C. (2022). Studying Alignment in a Collaborative Learning Activity via Automatic Methods: The Link Between What We Say and Do. *Dialogue & Discourse*, *13*(2), 1-48.

Ouyang, F., & Jiao, P. (2021). Artificial intelligence in education: The three paradigms. *Computers and Education: Artificial Intelligence*, *2*, 100020.

Ouyang, L., Wu, J., Jiang, X., Almeida, D., Wainwright, C., Mishkin, P., ... & Lowe, R. (2022). Training language models to follow instructions with human feedback. Advances in neural information processing systems, 35, 27730-27744.

Panadero, E. (2017). A review of self-regulated learning: Six models and four directions for research. *Frontiers in psychology*, *8*, 422.

Perkins, D. N., Jay, E., & Tishman, S. (1993). Beyond abilities: A dispositional theory of thinking. *Merrill-Palmer Quarterly (1982-)*, 1-21.

Pintrich, P. R. (2000). The role of goal orientation in self-regulated learning. In *Handbook of self-regulation* (pp. 451-502). Academic Press.

Rogat, T. K., & Linnenbrink-Garcia, L. (2011). Socially shared regulation in collaborative groups: An analysis of the interplay between quality of social regulation and group processes. *Cognition and instruction, 29*(4), 375-415.

Shaffer, D. W., Collier, W., & Ruis, A. R. (2016). A tutorial on epistemic network analysis: Analyzing the structure of connections in cognitive, social, and interaction data. *Journal of learning analytics*, *3*(3), 9-45.

Sitzmann, T., & Ely, K. (2011). A meta-analysis of self-regulated learning in work-related training and educational attainment: what we know and where we need to go. *Psychological bulletin, 137*(3), 421.

Tong, D., Jin, B., Tao, Y., Ren, H., Atiquil Islam, A. Y. M., & Bao, L. (2025). Exploring the role of human-AI collaboration in solving scientific problems. *Physical Review Physics Education Research, 21*(1), 010149.

Van Der Aalst, W. M. (2021). Hybrid Intelligence: to automate or not to automate, that is the question. *International Journal of Information Systems and Project Management*, *9*(2), 5-20.

Vauras, M., Iiskala, T., Kajamies, A., Kinnunen, R., & Lehtinen, E. (2003). Shared-regulation and motivation of collaborating peers: A case analysis. *Psychologia, 46*(1), 19-37.

Venkatesh, V., Thong, J. Y., & Xu, X. (2012). Consumer acceptance and use of information technology: extending the unified theory of acceptance and use of technology. *MIS quarterly*, 157-178.

Wang, Y., Gong, S., Cao, Y., & Liu, Y. (2025). Parallel empathy or reactive empathy? The role of emotional support provided by affective pedagogical agent in online learning. *The Internet and Higher Education*, 101035.

Wang, H., Wang, C., Chen, Z., Liu, F., Bao, C., & Xu, X. (2025). Impact of AI-agent-supported collaborative learning on the learning outcomes of University programming courses. *Education and Information Technologies*, 1-33.

Wegerif, R. (2013). *Dialogic: Education for the Internet age*. Routledge.

Wu, R., & Yu, Z. (2024). Do AI chatbots improve students learning outcomes? Evidence from a meta-analysis. *British Journal of Educational Technology*, *55*(1), 10-33.

Zhai, X., Nyaaba, M., & Ma, W. (2025). Can generative AI and ChatGPT outperform humans on cognitive-demanding problem-solving tasks in science?. *Science & Education*, 34(2), 649-670.

Zhu, Y., Liu, Q., & Zhao, L. (2025). Exploring the impact of generative artificial intelligence on students' learning outcomes: A meta-analysis. *Education and Information Technologies*, 1-29.

Zhu, G., Sudarshan, V., Kow, J. F., & Ong, Y. S. (2024, June). Human-generative AI collaborative problem solving who leads and how students perceive the interactions. In *2024 IEEE Conference on Artificial Intelligence (CAI)* (pp. 680-686). IEEE.

Zimmerman, B. J. (2000). Attaining self-regulation: A social cognitive perspective. In *Handbook of self-regulation* (pp. 13-39). Academic press.

# Appendix A. Task description

***Task A.*** "Niche academic disciplines" refer to fields that are highly specialized or niche within the academic system and receive relatively limited public attention, yet contain distinctive value and untapped potential. These disciplines may carry a deep historical and cultural legacy, or hold important significance for scientific and technological development and human progress. In the era of general-purpose artificial intelligence, as technology advances and knowledge systems evolve, some niche academic disciplines may gain renewed vitality, but they will also face a range of unique challenges. Choose one niche academic discipline that you believe has strong potential for future development. Analyze why it may be revitalized in the AI era, how AI technologies could support the discipline's preservation, innovation, and societal application, and what challenges it may encounter in its future development.

***Task B.*** Many science fiction novels and films depict scenarios in which artificial intelligence destroys humanity. Imagine a future "AI terrorist" that is skilled at exploiting cutting-edge technologies to carry out attacks. If you were an agent in a security agency, analyze the major security risks such an AI terrorist could pose, and identify the domains or technological layers where these risks might arise. In response, how would you prevent and monitor these threats, and what technologies would you use to design effective countermeasures? Drawing on specific scenarios and the underlying technical principles, propose solutions that are both feasible and innovative, and discuss the future development of AI security and defense technologies.